# Determining Electron Beam Lateral Coherence in a Scanning Electron Microscope Using Electron Diffraction

Evelijn Akerboom[a*], Fatemeh Kiani[b], Giulia Tagliabue[b], Wiebke Albrecht[a], Joanne Etheridge[c], F. Javier García de Abajo[d,e], and Albert Polman,[a*]

[a]Center for Nanophotonics, NWO-Institute AMOLF, Science Park 104, 1098 XG Amsterdam, the Netherlands

[b] Laboratory of Nanoscience for Energy Technologies (LNET), STI, École Polytechnique Fédérale de Lausanne, 1015 Lausanne, Switzerland

[c] School of Physics and Astronomy and Monash Centre for Electron Microscopy, Monash University, Victoria 3800, Australia

[d] ICFO-Institut de Ciencies Fotoniques, The Barcelona Institute of Science and Technology, 08860 Castelldefels, Barcelona, Spain

[e] ICREA-Institució Catalana de Recerca i Estudis Avançats, Passeig Lluís Companys 23, 08010 Barcelona, Spain

* Corresponding email: e.akerboom@amolf.nl, a.polman@amolf.nl

---

We develop and characterize scanning transmission electron microscopy (STEM) capabilities within a scanning electron microscope (SEM) to investigate the effective lateral coherence of the electron beam (e-beam) in the specimen plane. Using single-crystalline Au flakes and a sample composed of a monolayer of graphene, we obtain high-quality selected-area electron diffraction (SAED) maps and convergent-beam electron diffraction (CBED) patterns, validating the system's ability to probe crystallographic information at an acceleration voltage of 30 keV. Building on these capabilities, we implement a method, which is adapted from techniques traditionally used in transmission electron microscopy, to measure the degree of lateral coherence of the e-beam in the specimen plane of the SEM. By analyzing interference between electrons with two different wave vectors separated by 0.031 Å$^{-1}$, we extract a lower limit for the degree of lateral coherence over 5% of the e-beam diameter of approximately 60%. These coherence values are sufficient to enable quantum-coherent electron-light-matter interaction experiments in the SEM.

# Introduction

The ability to perform quantum-coherent electron-light-matter interaction experiments in a scanning electron microscope (SEM) has recently attracted increasing interest. Compared to transmission electron microscopes (TEMs), SEMs operate at substantially lower electron energies, typically between 1 and 30 keV, enabling access to stronger interaction regimes that are difficult to reach at higher electron energies. In particular, slower electrons have an increased interaction strength with optical near fields[1], making SEMs an attractive platform to study electron-photon coupling. Recently, approaches have emerged aimed at probing the quantum nature of electron photon interactions[2], such as photon-induced near-field electron microscopy[3]. A key for such experiments is a detailed understanding of the coherence properties of the electron beam (e-beam) at the specimen plane.

The coherence of the e-beam can be divided into temporal and lateral coherence. Temporal coherence is determined by the energy spread of the e-beam. For a Schottky FEG, the energy spread is approximately 0.3-0.8 eV, corresponding to a temporal coherence length at the sample plane of 1000-500 nm[4]. The lateral coherence is related to the effective size of the electron source and the probe-forming electron optics and aperture configuration. For a Schottky FEG, the current within the lateral coherent area is in the order of 100 pA[5]. Only a few physical processes, such as Coulomb interactions between electrons, can significantly alter the lateral coherence. Quantitative knowledge of the lateral coherence is essential for evaluating the feasibility of quantum-coherent electron-light-matter interactions and diffraction in the SEM. However, experimental methods to measure lateral coherence in SEMs are not well established.

In TEMs, several established methods exist to assess the lateral coherence or effective source size of the electron probe[6–13], including direct imaging of the source[6], direct measurements of Airy diffraction from a selective aperture[8,9], and holographic methods[13,14]. Many of these methods require long camera lengths or additional electron-optical elements, such as a biprism, which are not accessible in conventional SEMs and cannot be directly transferred. An alternative method for assessing lateral coherence involves recording a Ronchigram through convergent-beam electron diffraction (CBED) where the diffraction disks partially overlap [6,11,12]. In TEMs, CBED has been used to study van der Waals materials[15–19]. In particular, Kazmierczak *et al.* used it to show strain in twisted bilayer graphene[20], while Latychevskaia *et al.* showed how to use a twisted bilayer of graphene and hBN to determine out-of-plane atomic displacements[17].

Until recently, implementing diffraction measurements in SEMs was not well established, because generally SEMs lack the electron detectors required to efficiently record diffraction patterns. However, four-dimensional scanning transmission electron microscopy (4D-STEM) is increasingly deployed in SEMs [21–29], where the e-beam is scanned over the specimen, and for every e-beam position, a two-dimensional (2D) diffraction pattern is collected using a 2D single-electron detector, resulting in a rich 4D dataset that contains both spatial and reciprocal-space information[30]. This development creates new opportunities for characterizing the coherence properties of low-energy e-beams and extending diffraction-based techniques to the SEM.

In this work, we develop 4D-STEM in the SEM and use it to determine the effective lateral coherence of a 30 keV e-beam. To test the measurement geometry, we first demonstrate selected-area electron diffraction (SAED) and map the crystal orientation of ultra-thin Au flakes. We then obtain CBED patterns of van der Waals materials. We use a monolayer of graphene to characterize the SEM's semi-convergence angle and then adapt a technique used in TEMs to measure the degree of lateral coherence using a twisted bilayer of graphene. The results demonstrate that 4D-STEM can be used to quantify coherence properties of SEMs and indicate that the observed coherence is sufficient for future studies of quantum-coherent electron-light-matter interactions in the SEM.

# Results

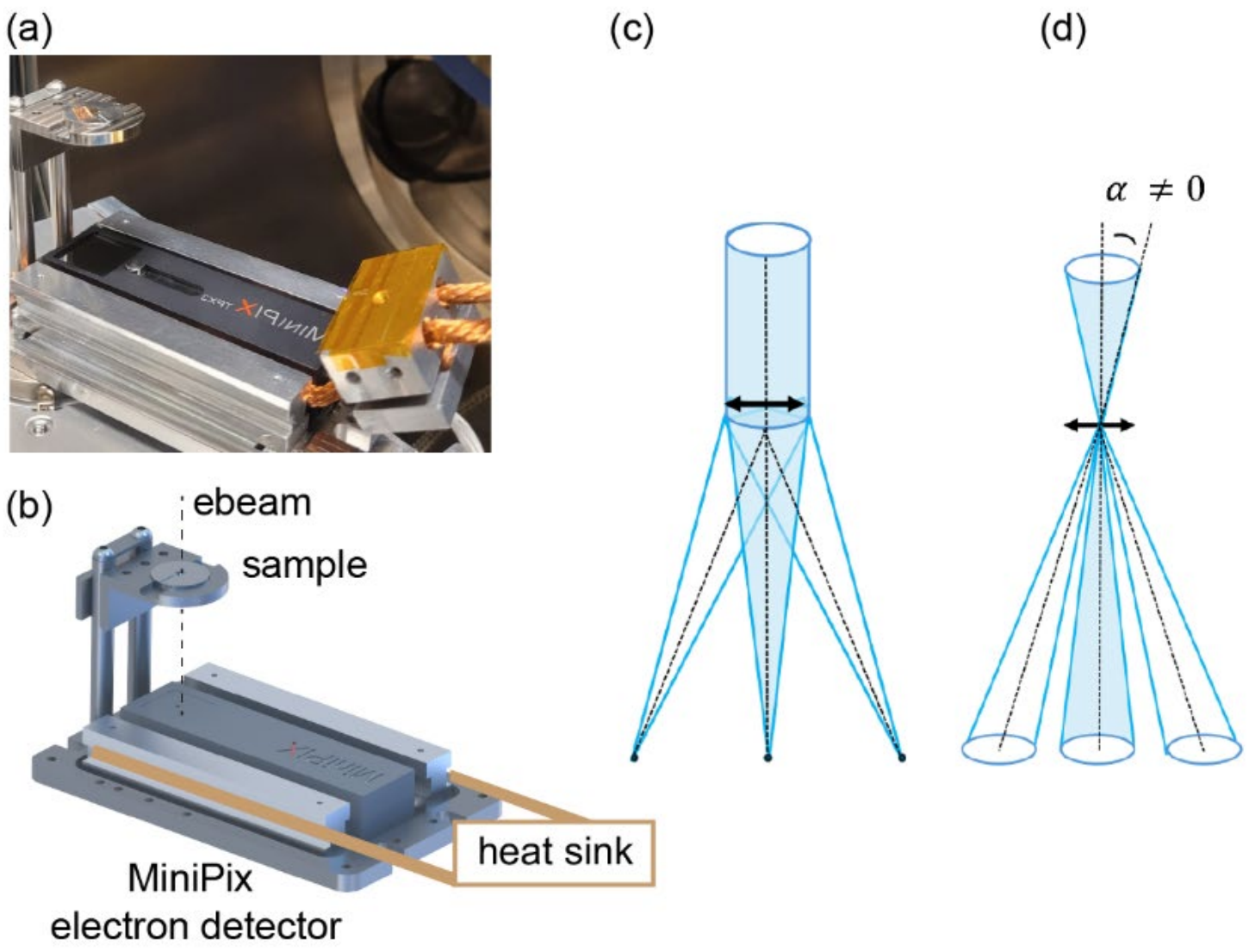

*Figure 1. Experimental setup for 4D-STEM in a SEM. (a) Photograph and (b) schematic representation of the STEM holder with the single-electron detector underneath. Schematic representation of two measurement geometries: (c) diffraction of a parallel e-beam, and (d) convergent beam electron (CBED) by a specimen (black arrow).*

To measure the lateral coherence of the e-beam in the SEM, we start with establishing 4D-STEM. We use a FEI Quanta 650 SEM equipped with a single-electron detector (MiniPIX TPX3, Advacam s.r.o) mounted below the sample (see Figure 1a and 1b for a photograph and a schematic). Depending on the lens settings, we achieve a semi-convergence angle between 1 and 8 mrad, which allows us to collect either SAED using a parallel beam (Figure 1c) or CBED patterns (Figure 1d). For a detailed description of the experimental setup, see Methods.

## Selected-area electron diffraction using Au flakes

We first perform 4D-STEM measurements on a single-crystalline Au flake fabricated using gap-assisted chemical synthesis[31]. The Au flakes are deposited on a $Si_3N_4$ membrane of 15 nm thickness, which is sufficiently thin to be transparent to the e-beam, even at the relative low kinetic energies used in the SEM. Measurements are conducted at an electron energy of 30 keV, a WD of 25 mm, and a camera length between the specimen and the electron detector of 4.7 cm. Figure 2a shows a secondary electron (SE) image of the Au flake above the $Si_3N_4$ membrane. The corresponding diffraction pattern recorded by summing over the total flake is shown in Figure 2b, where the background diffraction through $Si_3N_4$ is subtracted. The hexagonal symmetry of the pattern confirms diffraction from the face-centered cubic (fcc) lattice of Au along the <111> zone axis. A schematic of the diffraction pattern geometry for the <111> zone axis is shown in Figure 2c. Due to the summing of the diffraction patterns over the entire flake, the impact of local structural information such as strain and sample curvature, is averaged out over the measured area, resulting in the mismatch between Figure 2b and 2c.

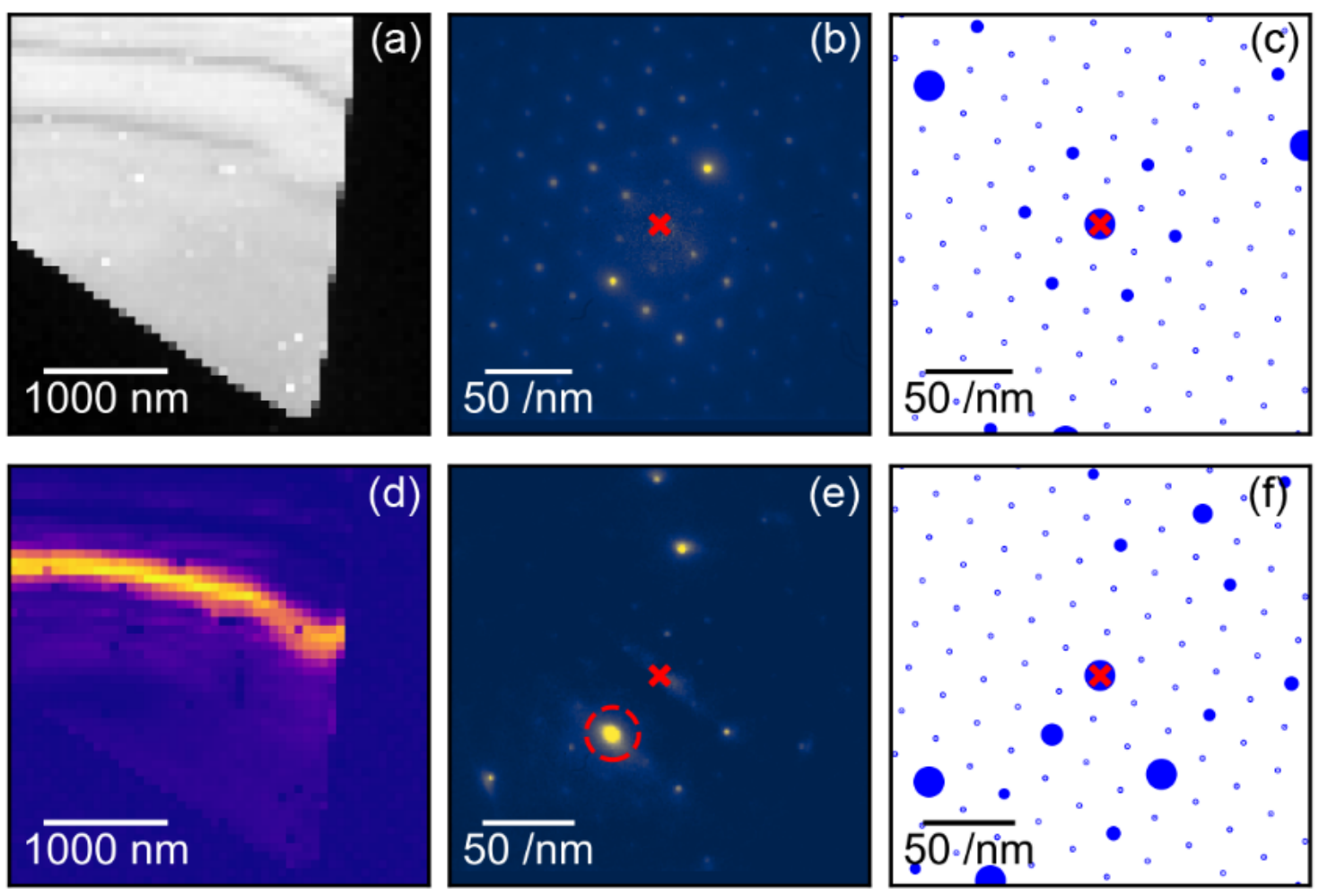

*Figure 2. Selected-area electron diffraction of a thin Au flake. (a) SE image of the thin Au flake. (b) Summed diffraction pattern collected over the entire flake, and (c) schematic of the diffraction pattern geometry for the <111> zone axis, where the size of each spot is indicative of the intensity of the diffraction spot. (d-f) Analysis of the sample orientation with respect to the e-beam. (d) Transmitted electron image generated using the reflection corresponding to the red dashed circle in (e) showing a bend contour. (e) Diffraction pattern for the position of the probe on the bend contour, where this spot has the maximum intensity (relative to other patterns in the 4D-STEM dataset), and (f) schematic of the diffraction pattern geometry for an incident e-beam direction at 6.5° relative to the <111> zone axis.*

The intensity variation between diffraction spots depends on the sample tilt relative to the e-beam direction, resulting in the two brighter spots observed on both sides of the center beam in Figure 2b. To investigate this further, we select one diffraction spot (marked by the red dashed circle) and record a transmitted electron image using only electrons from this diffraction order, revealing a bend contour, as shown in Figure 2d. This image reveals spatial regions within the flake where the selected diffracted beam has high intensity. The corresponding diffraction pattern is shown in Figure 2e. A schematic corresponding to the geometry of this diffraction pattern is shown in Figure 2f, where a tilt angle of 6.5° relative to the <111> zone axis is assumed. Such bend contours are not unexpected for such thin Au flakes. In both experiment and the schematic, we see the typical Laue circle where the Ewald sphere intersects the reciprocal lattice points with its center on the top left of the image. We note that, due to the lower electron energy compared to regular TEM, the Ewald sphere has a larger curvature.

## Convergent-beam electron diffraction of graphene

Next, we adjust the objective lens current and decrease the WD to enter a regime where we have a large semi-convergence angle (approximately 7.5 mrad, depending on the aperture settings). This allows us to study CBED patterns, which form the basis for the lateral coherence measurement that follow. As a model system, we first study a sample composed of a monolayer of graphene supported by a lacey carbon grid. Figure 3a shows the SE image of the sample, where the area in between the grid is composed of a monolayer of graphene. Using an electron energy of 30 keV, a WD of 4 mm, and a camera length of 6.65 cm, we record the CBED pattern shown in Figure 3b. We clearly see the diffraction pattern associated with the hexagonal crystal lattice of graphene. However, instead of recording discrete diffraction spots as before, we record diffraction disks due to the range of incident angles of the electron probe. When we calculate the expected CBED pattern geometry from the known camera length and semi-convergence angle and overlay the disks with the data (white dashed contours), we find excellent agreement. Small deviations in the disk positions may arise from local strain within the graphene layer. However, with our current angular resolution of ~1 mrad, such fine variations cannot be resolved.

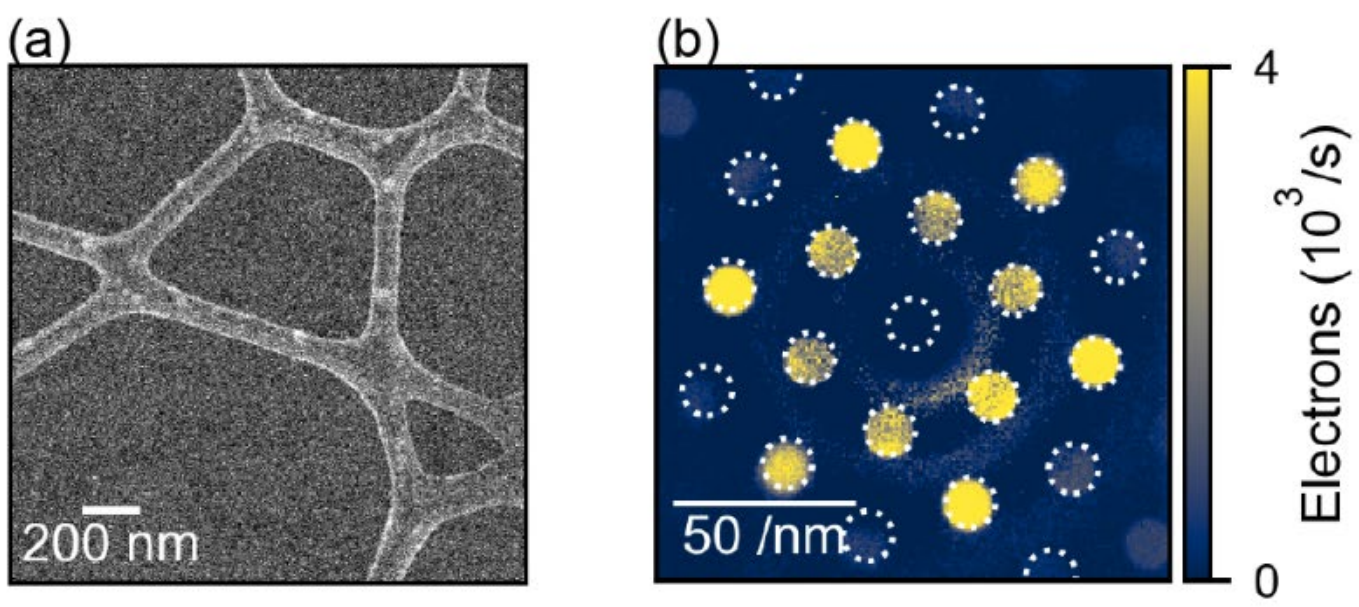

*Figure 3. CBED of monolayer graphene. (a) SE image of a monolayer of graphene supported by lacey carbon and (b) CBED pattern from this monolayer using an electron energy of 30 keV, a working distance of 4 mm, a camera length of 6.65 cm, and a semi-convergence angle of 7.5 mrad.*

## Lateral coherence analysis using twisted bilayer of graphene

Having established the CBED geometry in the SEM, we now apply it to measure the lateral coherence of the e-beam, using a method adapted from TEM [12]. To determine the lateral coherence, we collect CBED patterns on a bilayer of graphene under conditions where the diffraction disks overlap. Figure 4 shows a schematic representation of the experiment. The detector is placed at a camera length $Z$ below the focal point of the e-beam, while the sample is placed a distance $\Delta f$ above the focal point. The bilayer sample consists of two graphene layers spaced by a vertical distance $d$ and a twist angle $\beta$, with each layer acting as a diffracting plane for the e-beam. In a bilayer of graphene, the interlayer spacing equals 3.35 Å [17], which is small compared to $\Delta f$ and can be neglected in the calculation. The relative rotation between the two layers introduces a lateral displacement of their reciprocal lattice points ($\boldsymbol{g_1}$ and $\boldsymbol{g_2}$) relative to each other, such that electrons diffracted by each layer that end up at the same detector position have slightly different transverse wave vector $\boldsymbol{k}$. This produces a sufficient phase difference between the diffracted beams to generate a visible interference pattern, as illustrated schematically in Figure 4. The visibility of this interference pattern is then related to the degree of lateral coherence of the e-beam at a specific spatial frequency set by the specimen. For a bilayer of graphene, the difference between the reciprocal lattice vectors of the two layers is the Moiré wave vector, given by[32]

$$|\boldsymbol{g_1} - \boldsymbol{g_2}| = 2|\boldsymbol{g_1}| \sin(\beta/2), \qquad 2$$

where $|\boldsymbol{g_1}|$ depends on the diffraction order and we can use several diffraction orders to probe multiple Moiré wave vectors within the same measurement.

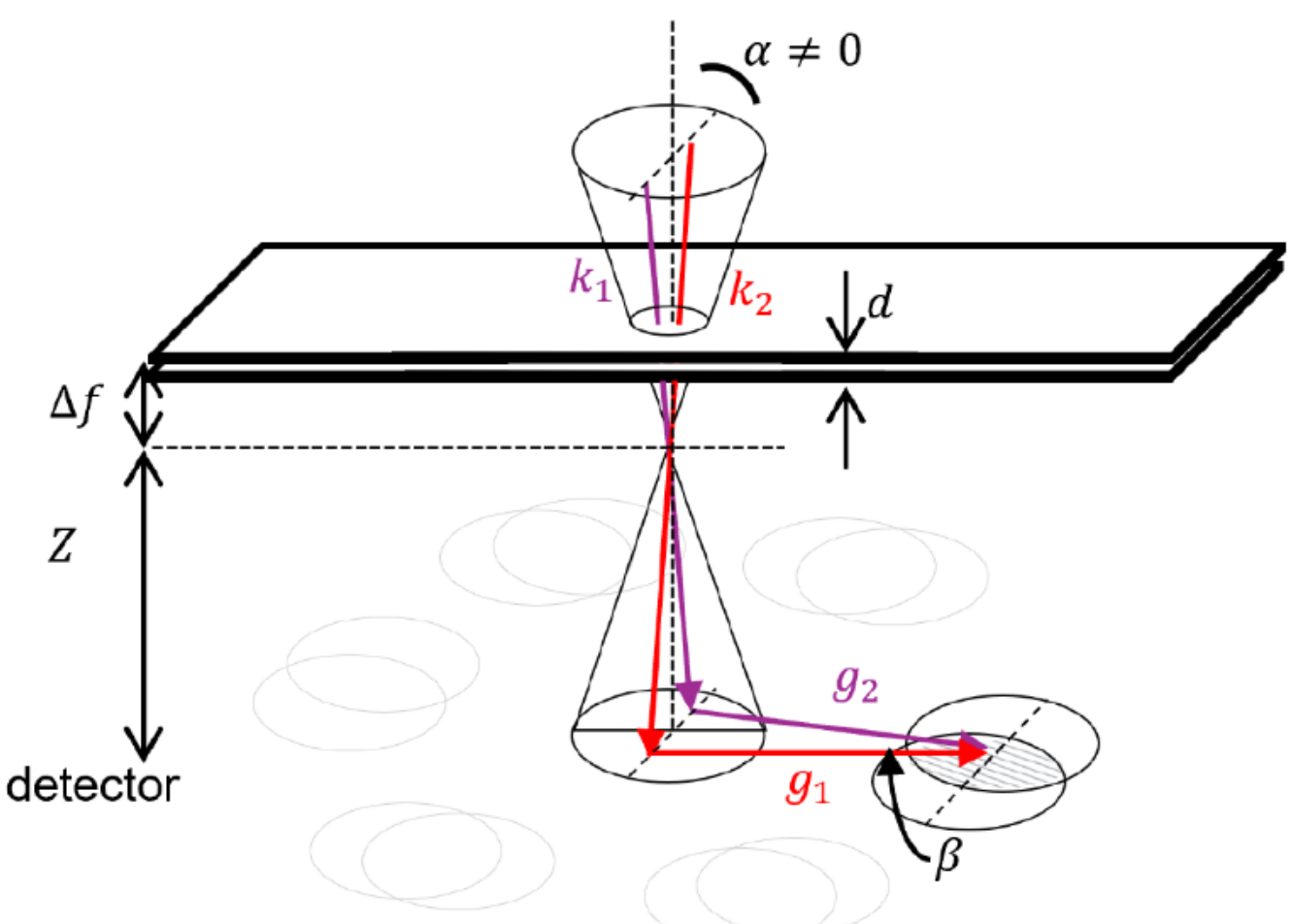

*Figure 4. Schematic representation of CBED for a twisted bilayer of a 2D material resulting in overlapping diffraction disks.*

### *Calculation*

First, we calculate the expected interference pattern for a perfectly laterally coherent beam. We start by determining the positions of the two virtual point sources, which depend on the twist angle and defocus height. The resulting interference intensity distribution is obtained from the coherent superposition of the two waves (see Supporting Information for a detailed description). For the calculation, we use a twist angle$\beta = 0.6°$, an electron energy of 30 keV, and a semi-convergence angle $\alpha = 7.5$ mrad. The sample is placed $\Delta f = 6\ \mu m$ above the focal plane and the camera length is $z = 6.65$ cm.

Figure 5a shows the corresponding calculated CBED pattern for a perfectly coherent beam. The overall diffraction geometry closely resembles that observed experimentally for a monolayer of graphene (Fig. 3b). However, within each diffraction disk, we now observe a distinct interference pattern. The period of the interference fringes is lower for the higher-order diffraction disks, and their orientation rotates according to the position of the disk with respect to the direct beam: the fringes are always perpendicular to the vector connecting the two disks. Because the twist angle is small, the two disks are nearly coincident and cannot be easily distinguished.

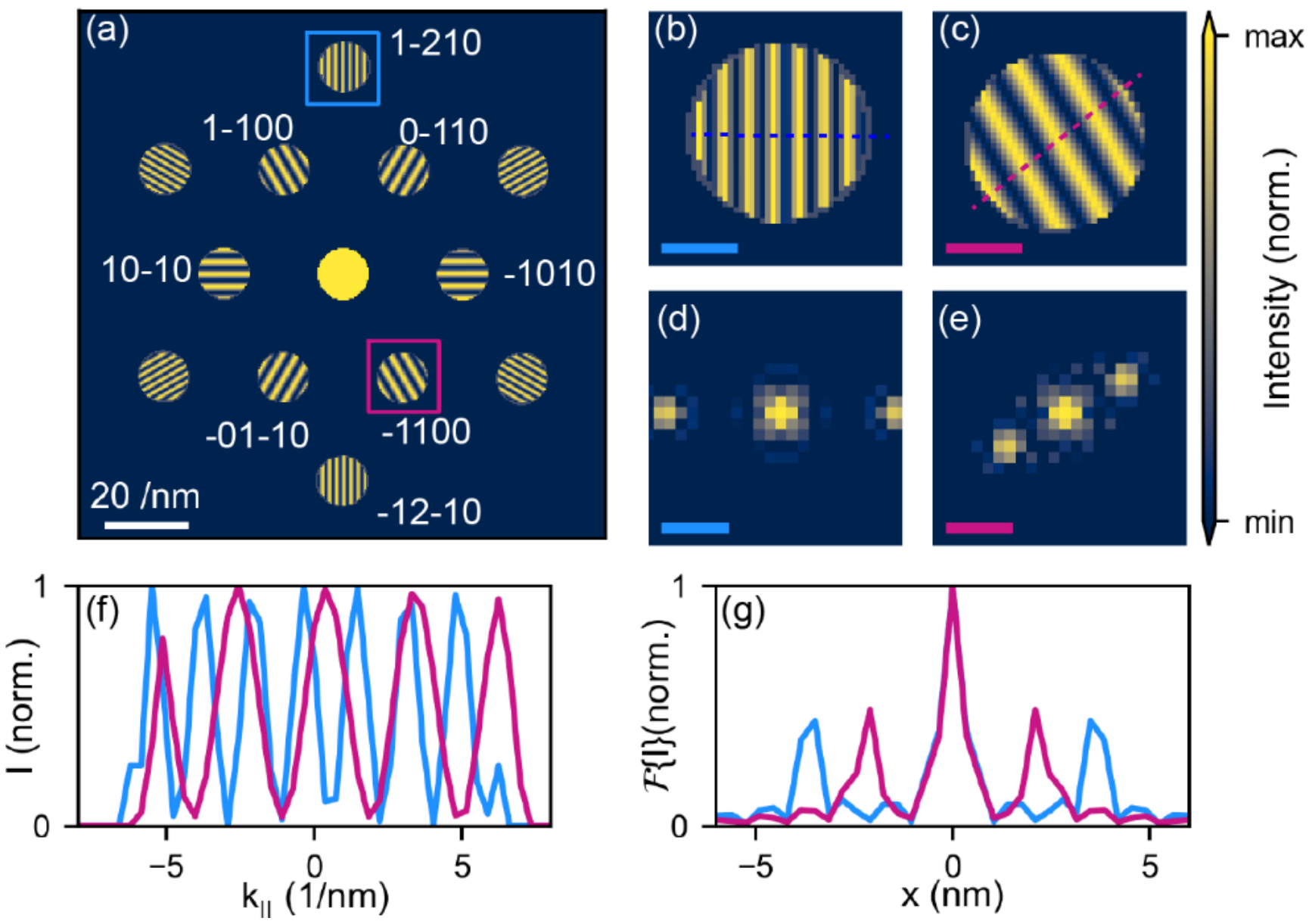

*Figure 5. Calculated CBED pattern for a twisted bilayer of graphene with crystal orientations indexed. (a) Full CBED pattern covering up to second-order diffraction, calculated for a twist angle $\beta = 0.6°$ and a defocus $\Delta f = 6\ \mu m$ (see Supporting Information). (b,c) Enlarged images from (a) for two diffraction disk orientations, (1-210) (blue) and (-1100) (purple) (scale bar 5 nm$^{-1}$) along with (d,e) their 2D Fourier transform (scale bar 2 nm), respectively. (f) Line scans perpendicular to the interference patterns in (b,c) in blue and purple, respectively, and (g) their Fourier transforms.*

The lateral coherence is probed at a wave vector given by the specimen (see Eq. 2), which depend on the twist angle and the diffraction order. For the chosen twist angle, the Moiré wave vectors are equal to 0.031 Å$^{-1}$ and 0.053 Å$^{-1}$ for the first and second diffraction order, respectively (see Supporting Information for a calculations)[32]. Therefore, we analyze the fringes in more detail for the (1-210) and the (-1100) diffraction disks, indicated by the blue and purple squares in Figure 5a. Enlarged images are shown in Figures 5b and 6c, respectively. Around the outer region of each disk, there is an area without overlap, such that no interference fringes are observed. Taking the Fourier transform of the enlarged images allows us to study the distance between the virtual sources contributing to the interference pattern, shown in Figure 5d,e with the scale bar indicating 2 nm. We show a line scan perpendicular to the interference fringes in Figure 5b,c (corresponding to the dashed line), and take the Fourier transform to study the visibility and spacing of the fringes, as shown in Figure 5f,g. The distance related to this interference is 2.2 nm and 3.5 nm for the first- and second-order diffraction spots, respectively. This distance is associated with the real-space distance between the virtual sources. Furthermore, since in the calculation we assume a perfectly coherent e-beam and an equal scattering intensity from each layer, and ignore effects of inelastic scattering, the fringe visibility reaches 100%.

### *Experiments*

To experimentally measure the lateral coherence in the SEM, we measure CBED patterns from a commercial bilayer of graphene (Ted Pella), consisting of many small flakes with varying twist angles and number of layers. We place the sample $\Delta f = 6\ \mu m$ above the focal plane, using a final aperture of 1000 $\mu$m and a WD of 4 mm, resulting in an e-beam semi-convergence angle $\alpha = 7.5$ mrad and e-beam diameter at the specimen plane of 80 nm. Figure 6 shows the CBED results of an area exhibiting clear interference fringes. Figure 6a shows the full CBED pattern recorded on the single-electron detector after background subtraction to remove contributions from inelastically scattered electrons. The dark central disk in the image is an artifact of the background subtraction, related to oversaturation of the detector in this area. In addition, within each diffraction disk, a dark spot appears at the same position, which we assign to a shading effect due to a small contamination that blocks part of the transmitted electrons.

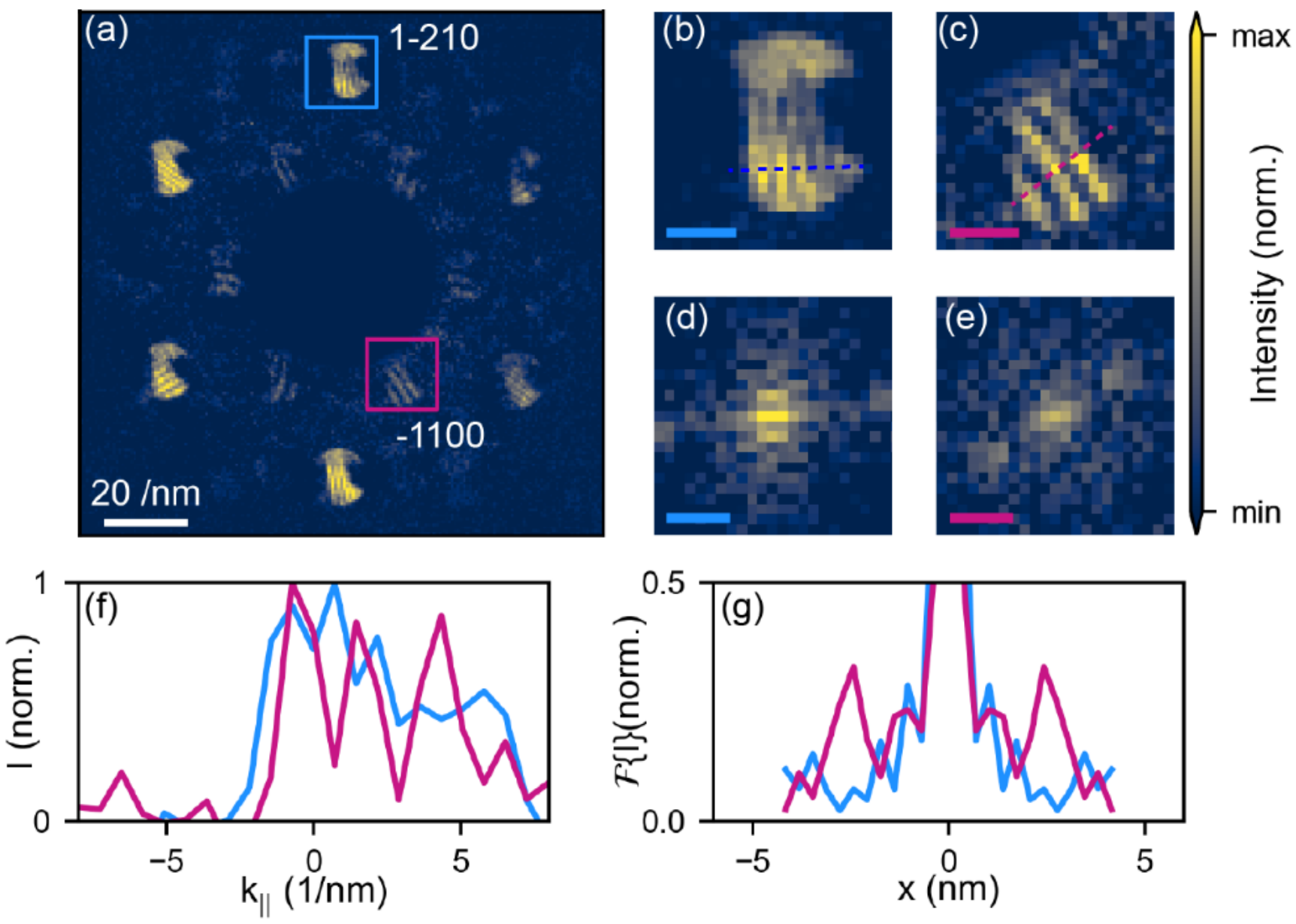

*Figure 6. Measured CBED pattern for a twisted bilayer of graphene, using a semi-convergence angle $\alpha = 7.5$ mrad. (a) Full CBED pattern with the background subtracted, resulting in a dark disk in the center. (b,c) Enlarged images from (a) for the (1-210) (blue) and (-1100) (purple) diffraction disks (scale bar 5 nm$^{-1}$) along with (d,e) their 2D Fourier transforms (scale bar 2 nm), respectively. (f) Line scans perpendicular to the interference fringes in (b,c) (in blue and purple, respectively), and (g) their Fourier transforms.*

Nonetheless, around the shaded region a distinct interference pattern is visible. The fringes rotate according to the orientation of the diffraction spot with respect to the central beam, as expected from the simulations (Figure 5a). This behavior is more clearly seen in the enlarged images of two specific spots indicated by the blue and purple squares shown in Figure 6b,c, respectively. The Fourier transforms of the interference patterns (Figures 6d,e) reveal components corresponding to the fringe spacing, although the image quality is affected by the shaded regions in the CBED disks. The fringe orientation and periodicity match the calculated results from Figure 5 well, demonstrating good agreement between experiments and calculation for $\Delta f = 6\ \mu m$ and $\beta = 0.6°$.

To measure the fringe visibility, we take line scans of Figure 6b,c, indicated by the blue and purple dashed lines, respectively, and compute the Fourier transforms, shown in Figure 6f,g. Although the visibility is substantially decreased compared to the calculation (note the different $y$ scale in panel (g)), we clearly see a peak in the Fourier transform. For the (-1100) diffraction disc (purple), we find a peak at 2.4 nm with a visibility of approximately 60%, while for the (1-210) diffraction disk (blue), a peak is found at 3.5 nm with a visibility of 20%. Thus, the measurement provides two data points on the lateral coherence curve of the SEM beam, as summarized in Table 1. We observe a decreasing trend in visibility for larger wave vectors, as expected given the limited coherence of the e-beam. When compared to the maximum accessible e-beam wave vector of 0.63 Å$^{-1}$, these measurements sample only part of the full range. A more complete reconstruction of the lateral coherence function and probe shape would require repeating the measurement for other twist angles and additional diffraction orders, thereby covering a broader wave vector separation range. However, the finite pixel size of the detector imposes a lower bound on the fringe period that can be resolved (see Supporting Information for details).

*Table 1. Wave vectors connecting the diffraction points of a twisted bilayer graphene sample for the first and second diffraction orders and a twist angle of 0.6°. We also give the corresponding visibility of the interference fringes from Figure 6.*

| Diffraction order | $\|\boldsymbol{g_1} - \boldsymbol{g_2}\|$ (Å$^{-1}$) | Visibility (%) |
|---|---|---|
| 1 | 0.031 | 60 |
| 2 | 0.053 | 20 |

To obtain insight into the spatial coherence on the specimen plane we divide the value for $|\boldsymbol{g_1} - \boldsymbol{g_2}|$ for the first order diffraction by the accessible e-beam wave vector (0.63 Å$^{-1}$) given by semi-convergent angle and find a value of 5%. This implies that for 5% of the beam diameter the beam is coherent to 60%. This is a lower limit as multiple electron scattering and inelastic scattering, specimen contamination, and the fact the exact number of atomic layers in each flake is unknown, will reduce fringe visibility.

# Conclusion

In this work, we demonstrated the development and application of CBED within a SEM operating at an electron energy of 30 keV to determine the degree of lateral coherence of the e-beam. We performed convergent-beam electron diffraction on graphene monolayers and observed the transition from discrete diffraction spots to diffraction disks, verifying the expected geometry through comparison with calculated CBED patterns. To assess the lateral coherence of the SEM's e-beam, we adopted a method inspired by TEM studies, analyzing overlapping diffraction disks from twisted bilayer graphene. Calculations for a perfectly coherent beam were compared with experimental data, from which we extracted a lower limit for the degree of lateral coherence of approximately 60% and 20% for a transverse electron wave vector of 0.031 Å$^{-1}$ and 0.053 Å$^{-1}$, respectively. This corresponds to a lower limit for the degree of coherence of 60% over 5% of the e-beam diameter.

Overall, this work constitutes a proof-of-concept of using CBED in the SEM to determine the degree of lateral coherence. This geometry takes advantage of the high signal-to-noise ratio for transmitted electrons and the different measurement capabilities enabled by low electron energies in SEMs. Using CBED, we characterize key e-beam parameters: the semi-convergence angle and the degree of lateral coherence. Looking ahead, 4D-STEM can be combined with cathodoluminescence spectroscopy to directly correlate structural and optical properties within a single experiment. Ultimately, the observed degree of lateral coherence enables single-electron/single-photon coincidence measurements to obtain further insight into the quantum properties of electron-light-matter interactions.

# Methods

## Experimental setup

All experiments are performed using a FEI Quanta 650 SEM with a thermally assisted field emission gun (FEG), where a tungsten tip is heated by a filament current while a strong extractor voltage pulls electrons from its apex. To perform 4D-STEM in the SEM, we use a single-electron counting detector (MiniPIX TPX3,

Advacam r.s.o) with a segmented silicon sensor that comprises 256x256 pixels (Timepix3 chip). The sample is loaded above the detector at a camera length that is varied between 1 and 7 cm. To stabilize the detector temperature during measurements, the holder is connected to a heat sink cooled with liquid nitrogen.

During 4D-STEM acquisition, the e-beam was raster-scanned over the sample and a 2D diffraction pattern for each position was recorded. Depending on the objective lens and aperture settings, we are either in a near-parallel-beam geometry, resulting in SAED patterns, or in a convergent-beam geometry, producing CBED disks. In each case, the WD was adjusted to keep the specimen in focus.

## Sample preparation

In this work, three samples were used. Thin single-crystalline Au flakes were used to validate transmission diffraction. The Au flakes were fabricated using gap-assisted chemical synthesis[31] and deposited on a $Si_3N_4$ membrane of 15 nm thickness (Ted Pella), which are sufficiently thin to transmit 30 keV electrons.

For the CBED measurements on graphene, commercial samples were used. A PELCO Single Layer graphene on lacey carbon (Ted Pella) was used to characterize the CBED geometry. For lateral coherence measurements, we used a PELCO 2-layer graphene supported by holey $Si_3N_4$ membrane (Ted Pella). These samples contained multiple graphene regions with different local twist angles and layer configurations. Regions showing overlapping diffraction disks and visible interference fringes were selected for analysis.

## Determination of spatial resolution and semi-convergence angle

We determined the spatial resolution of the SEM using an acceleration voltage of 30 keV and a calibration sample that consists of semi-spherical Au particles with diameters between 2 and 30 nm (see Figure 7a). Using an analytical method based on the power spectrum (2D Fourier transform analysis) [33,34], we find a spatial resolution of 4.4 nm at a WD of 3.8 mm using an acceleration voltage of 30 keV, with spot setting of 3.9 and a final aperture of 1000 $\mu$m. To determine the semi-convergence angle of the e-beam, we focus the e-beam on the sample plane, removing the sample and recording the direct beam in reciprocal space, using a camera length of 6.65 cm. The resulting e-beam image at the detector is shown in Figure 7b, corresponding to a semi-convergence angle of 8.4 mrad.

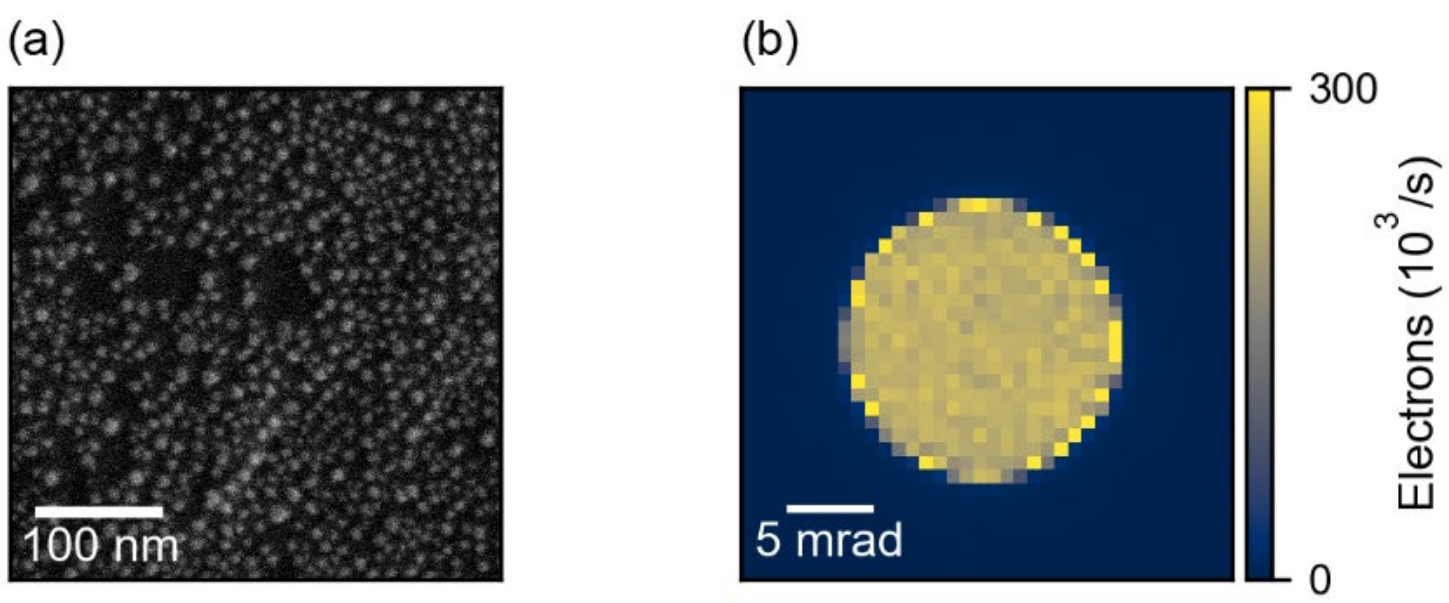

*Figure 7. SEM characterization. (a) High-resolution SE image of a calibration sample consisting of Au particles with diameters between 2 and 30 nm, and (b) focused e-beam in absence of a specimen recorded in reciprocal space on the single-electron detector at a camera length of 6.65 cm, showing a semi-convergence angle of 8.4 mrad.*

## SAED and CBED analysis

For the SAED measurements on the Au flakes, the SEM was operated at 30 keV, a e-beam spot setting of 3.9 and a final aperture of 30 μm, resulting in an e-beam current of 300 pA. The specimen was at a WD of 25 mm, at which the semi-convergence angle was measured to be approximately 1 mrad. The exposure time of the SAED pattern was 0.002 s, using an energy threshold of the electron detector of 4 keV and a bias voltage of 200 V. For the diffraction pattern averaged over large area, the SAED was averaged over the entire flake and the background diffraction from $Si_3N_4$ was subtracted.

To perform CBED measurements, the specimen was placed at a WD of 4 mm, using e-beam acceleration voltage of 30 keV, a spot setting of 3.9 and the final aperture of 1000 μm, resulting in a semi-convergent angle of 7.5 mrad and an e-beam current of 1 nA. The single electron detector was used with a threshold of 4 keV, a bias voltage of 200 V and an exposure time of 0.008 s. The measured CBED patterns were averaged over three measurements, and the radial median was subtracted to remove the signal from the direct beam. To perform the lateral coherence analysis, two diffraction disks were selected, and a line scan was taken over the center of the disk, perpendicular to the interference fringes. The Fourier transform is taken, and the visibility is found to be the ratio of the side peaks and the center peak.

## Supporting Information

The Supporting Information includes a calculation for convergent-beam electron diffraction from a bilayer of graphene with the calculated limitations of this measurement scheme.

## Funding Information

This work is financed by the Dutch Research Council (NWO) and has received funding from the European Research Council (ERC) under the European Union's Horizon 2020 Research and Innovation Program under Grant Agreements Nos. 101019932 (QEWS) and 101141220 (QUEFES). J.E acknowledges funding from the Australian Research Council Laureate Fellowship grant FL220100202. F.K and G.T acknowledge the Swiss National Science Foundation through the Eccellenza Grant 194181.

-------------- Appendix ----------------

# Determining Electron Beam Lateral Coherence in a Scanning Electron Microscope Using Electron Diffraction

Evelijn Akerboom[a*], Fatemeh Kiani[b], Giulia Tagliabue[b], Wiebke Albrecht[a], Joanne Etheridge[c], F. Javier García de Abajo[d,e], and Albert Polman,[a*]

[a]Center for Nanophotonics, NWO-Institute AMOLF, Science Park 104, 1098 XG Amsterdam, the Netherlands

[b] Laboratory of Nanoscience for Energy Technologies (LNET), STI, École Polytechnique Fédérale de Lausanne, 1015 Lausanne, Switzerland

[c] School of Physics and Astronomy and Monash Centre for Electron Microscopy, Monash University, Victoria 3800, Australia

[d] ICFO-Institut de Ciencies Fotoniques, The Barcelona Institute of Science and Technology, 08860 Castelldefels, Barcelona, Spain

[e] ICREA-Institució Catalana de Recerca i Estudis Avançats, Passeig Lluís Companys 23, 08010 Barcelona, Spain

* Corresponding email: e.akerboom@amolf.nl, a.polman@amolf.nl

---

## Convergent-beam electron diffraction calculation for graphene

In the convergent-beam electron diffraction (CBED) configuration sketched in Figure 6, the electron beam (e-beam) is interacting with a graphene bilayer with a distance *d* between the layers and a twist angle $\beta$. As a result, we observe two diffraction disk patterns on the detector, originating from electrons scattered from the first and second layers. Within the area where the diffraction disks overlap, electrons can come from either one of the layers. Therefore, we observe an interference pattern that resembles the superposition of electron waves originating from two virtual sources. The orientation of the interference fringes is always perpendicular to the vector connecting the centers of the two overlapping disks.

A quantitative analysis of these interference fringes would require a full multi-slice calculation[1], incorporating propagation over the distance *d* of the wavefield to the second layer, after scattering from the first layer. However, in this section, our aim is to obtain a qualitative understanding of the geometry of the interference patterns. To calculate the expected interference period, we derive the distance between the two sources. In two layers that are stacked on top of each other without translation or rotation, the distance between the two virtual sources is[2,3]

$$\Delta\rho = d\tan(\theta_i) \tag{1}$$

where $\theta_i$ is the i$^{th}$ order Bragg diffraction angle for graphene related to the lattice constant ($a = 2.46$ Å), and $d$ the distance between the layers ($d = 3.35$ Å). The distance between the virtual sources without twist angle is not large enough to create an interference pattern that is visible with the resolution of our detector[2]. The location of the virtual sources with a twist angle and defocus height $\Delta f$ is given by (see Figure 5b for a schematic)

$$\rho_1 = (\Delta f + d)\tan(\theta_i)\,\hat{x} \tag{2}$$

$$\rho_2 = \Delta f\tan(\theta_i)\,(\sin(\beta)\,\hat{x} + \cos(\beta)\,\hat{y}). \tag{3}$$

Note that we assume that the phase associated with propagation from the first layer to the second is unchanged, which is a reasonable approximation under the paraxial conditions here considered. The resulting interference pattern at the position $\boldsymbol{R} = (x, y, z)$ in the detector can be calculated by summing up electron waves originating from $S^1$ and $S^2$. We obtain

$$I(\boldsymbol{R}) \propto \left| e^{ik_0|\boldsymbol{R}-\rho_1|} e^{-i\pi\frac{|\rho_1|^2}{\lambda_e \Delta f}} + e^{ik_0|\boldsymbol{R}-\rho_2|} e^{-i\pi\frac{|\rho_2|^2}{\lambda_e(\Delta f + d)}} \right|^2. \tag{4}$$

The resulting interference pattern depends on the under- or overfocus, the interlayer distance, and the twist angle. The larger the overfocus, underfocus or the twist angle, the larger the spacing between the virtual sources, hence the smaller the interference period. If we consider a general case, where each layer of the bilayer could comprise more than a single monolayer of atoms, then, within the above approximations, the visibility of the fringes will be determined by both the ratio between the number of atomic layers in each layer within the twisted bilayer, and the lateral coherence of the e-beam (ignoring multiple scattering [1]). For an equal number of layers, lateral coherence is given by

$$\boldsymbol{\nu} = \frac{I_{max} - I_{min}}{I_{max}} \qquad 5$$

In our experiments, the visibility is approximately 20-60%. Because sample contamination, multiple scattering and variation in the number of atomic layers contributing to the diffraction spots can only reduce the observed visibility, this value represents a lower bound for the degree of lateral coherence at the spatial frequency set by the specimen. For a bilayer of graphene, the difference between the reciprocal lattice vectors of the two atomic monolayers is given by[4]

$$q_1(\beta) = 2|\boldsymbol{b}_i| \sin(0.5\beta), \qquad 6$$

where $\boldsymbol{b_i}$ is the magnitude of the reciprocal lattice vector corresponding to diffraction order $i$. The spatial frequencies associated with different twist angles are shown in Figure S1 for both the first (purple) and second (blue) diffraction orders. To probe the coherence function across the entire beam, we would need to measure twisted bilayer graphene with a range of twist angles. However, at some twist angles, the diffraction disks in our SEM geometry no longer overlap, making the interference measurement impossible with the current beam configuration. The blue shaded region indicates the accessible transverse wave vectors for which the diffraction disks overlap within the e-beam for a semi-convergence angle of 7 mrad.

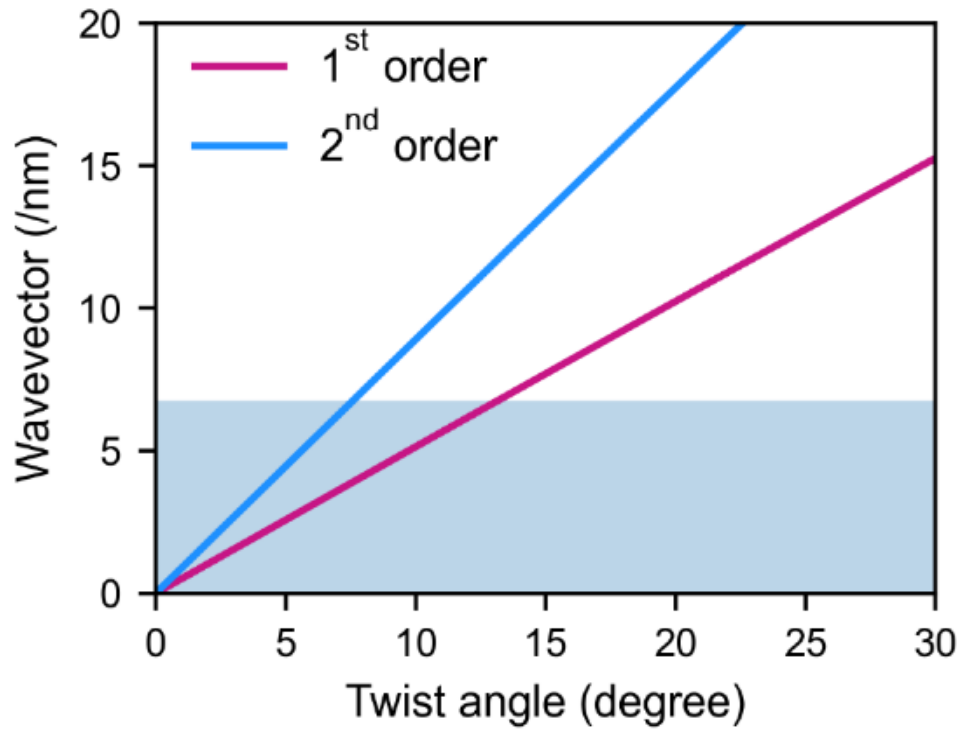

*Figure S 1. Wave vector associated with twisted bilayer of graphene. The solid lines show the relation between the twist angle and the associated wave vector for the first (purple) and second (blue) order diffraction disks of a bilayer of graphene. The accessible wave vectors in the e-beam with a semi-convergent angle of 7 mrad are shown for comparison (light blue).*

A second limitation of this method for determining the lateral coherence function is the finite pixel resolution of the single-electron detector coupled with the finite effective camera length. Because of the limited pixel size, the smallest interference period we can reliably distinguish corresponds to a virtual-source separation of 4 nm. Since the interference period depends on both the twist angle and the defocus height above the focal plane, we retain some flexibility in choosing these parameters. Figures S2a and S2b show the calculated virtual-source separations for the first and second diffraction orders, respectively. The dashed line marks the 4 nm resolution limit. Features to the right of this line cannot be resolved with our detector configuration. From these plots, we conclude that accessing higher spatial frequencies requires using larger twist angles while simultaneously reducing the defocus height to values on the order of only ~1 $\mu m$. Repeating the measurement under these conditions would extend the measurable range of wavevectors representing the lateral coherence distribution of the e-beam.

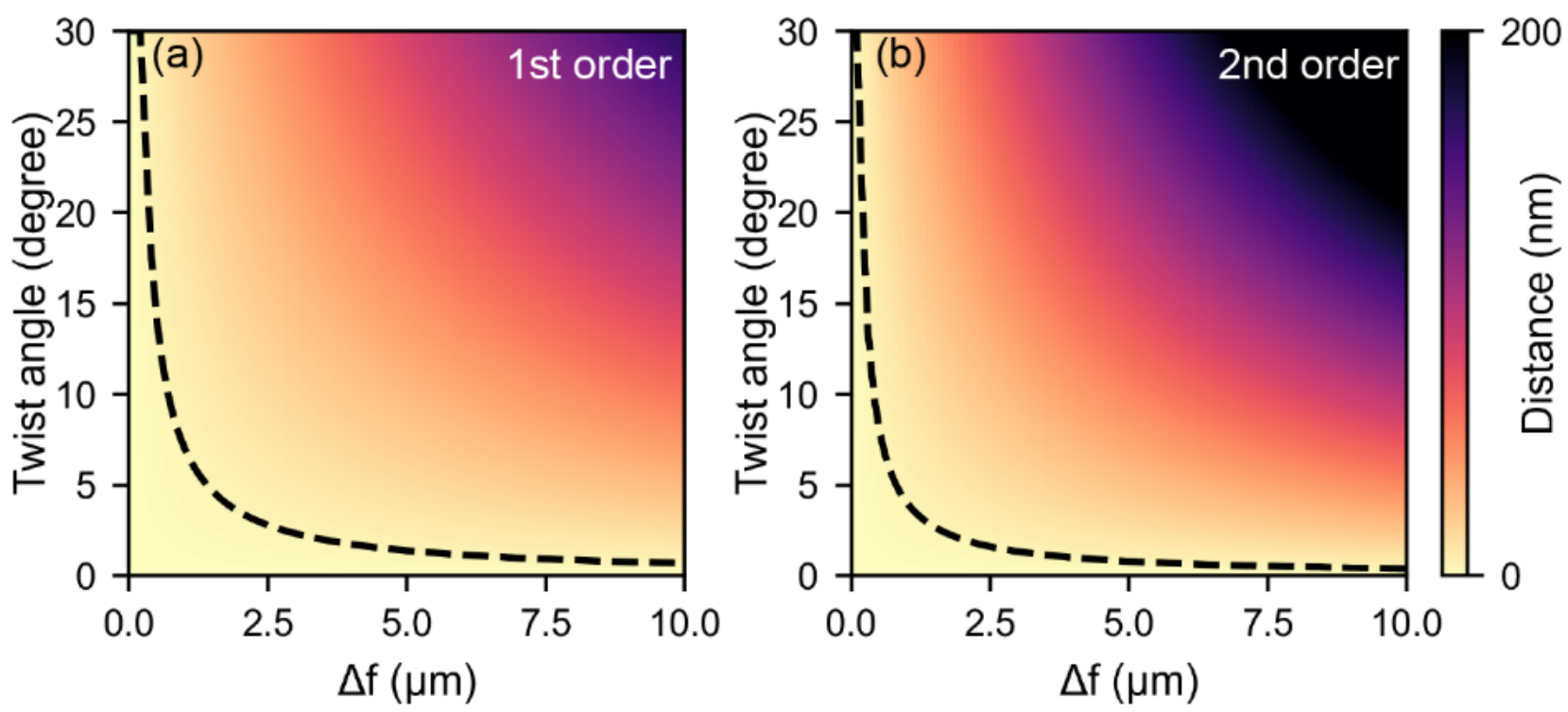

*Figure S 2. Limitation of measurements with the single-electron detector and electron-optical conditions available here. Calculated distance (colour scale) between the virtual sources for (a) the first and (b) the second order diffraction disks depending on the twist angle $\beta$ and defocus height $\Delta f$. The black dashed line shows the resolution limit of 4 nm. Only features on the left of this line can be resolved by the detector used here.*